B
B
\documentstyle[preprint,prd,aps]{revtex}
\global\arraycolsep=2pt

\begin{document}
\draft\pagenumbering{roma}

\author{Liu Yu-Xi$^{a,b}$\footnote{email: liuyx@itp.ac.cn}, Cao Chang-Qi $^b$}
\address{$^a$  Department of Physics, Peking University , Beijing 100871
China}
\address{$^b$ Institute of Theoretical Physics, Academia Sinica, 
P.O.Box 2735, Beijing 100080, China}
\title{Squeezing spectra of the output field for high density exciton laser}
\maketitle
\thispagestyle{empty}
\vspace{15mm}
\begin{abstract}
 The effect of the non-linear interaction between the high 
 density Wannier excitons is analysed. We use the Fokker-Planck equation in
 the positive P presentation and the corresponding 
  stochastic differential equation to study the composite system of a single
  mode cavity field and exciton under classical field pumping. 
 The small fluctuation approximation is made to get the 
  quadrature squeezing spectra of the output light field.
 The conditions for the squeezing of the either  quadrature component of 
 the output light are given.
\end{abstract}
\vspace{4mm}
\vspace{3.cm}
\newpage
\pagenumbering{arabic}
\begin{center}{1.Introduction}\end{center}
Recently, the excitons in semiconductor have been the subject of the intensive 
theoretical and the experimental investigation. One of reasons is  its
technological impact on  quantum electronics and photonics. Especially, people 
have studied some optical properties of the  excitons in the confined quantum 
system , such as  the quantum dot, the quantum wires, and the quantum 
well~\cite{a,b,c,d,e,f,g,h}.

It is well known that if the emitted light field is in squeezing state, 
 the fluctuation of one of its quadrature phase components  will be 
reduced at the expense of an enlargement of the fluctuation in the conjugate 
partner.  The quadrature phase component with reduced fluctuation could be 
applied to the low noise optical communications and the ultra-precise 
measurements. Many efforts 
have been dedicated to generate the squeezing 
state of the light field both theoretically and experimentally
~\cite{1,2,3,4,5,6,7} .

In this paper, the emission of the high density excitons in  quantum well 
under classical pump field is  studied. We assume
the quantum well is placed in the microcavity. The excitons interact  only with
a single mode of the cavity field. The reservoirs for both the excitons and 
cavity field have been taken into account. The stationary analysis of the 
system have been carried out by using the Fokker-Planck equation. The small 
fluctuations approximation around stationary state is used to studying the 
quadrature squeezing properties of the output light field. 

\begin{center}{2. Model and master equation}\end{center}
 We know when the density of the excitons become higher, the ideal bosonic model
  of the exciton is no longer adequate (In the case of GaAs Quantum 
 well, the ideal bosonic model become inadequate  when the density of the 
 excitons exceeds  $1.3\times10^{9}{\rm cm}^{-2}$~\cite{A}). One way of dealing
  with these deviations is to introduce an effective interaction between the 
  hypothetical ideal bosons.  We consider
the GaAs quantum well is placed in a microcavity, and the excitons
in the quantum well are pumped by the classical light with the frequency 
$\omega_{b}$. Under rotating wave approximation, we have the Hamiltonian as 
following:
\begin{equation}
\hat{H}=\hat{H}_{0}+\hat{H}_{R_{1}}+\hat{H}_{R_{2}}
+\hbar\hat{a}\Gamma^{+}_{1}+\hbar\hat{a}^{+}\Gamma_{1}
+\hbar\hat{b}\Gamma^{+}_{2}+\hbar\hat{b}^{+}\Gamma_{2},
\end{equation}
where:
\begin{eqnarray}
\hat{H}_{0}&=& \hbar \omega_{c}\hat{a}^{+}\hat{a}+
\hbar \omega_{b}\hat{b}^{+}\hat{b}+
\hbar G \hat{b}^{+}\hat{b}^{+}\hat{b}\hat{b} \nonumber  \\
& + &i\hbar g(\hat{a}^{+}\hat{b}-\hat{a}\hat{b}^{+})
+i\hbar (E\hat{b}^{+} e^{-i\omega_{b}t}-E^{*}\hat{b} e^{i \omega_{b}t}),
\end{eqnarray}
$\omega_{c}$ and $\omega_{b}$ are frequencies of the cavity field and the 
excitons respectively, $g$ is the coupling constant of the excitons and the 
cavity field, we assume it is a real number. The fifth term of the equation (2) 
is  the pumping term. The pumping field is taken  as  a classical driving 
field. 
$E$ is in proportion to its amplitude. 
$\hat{a}^{+}(\hat{a})$ is creation (annihilation)
operator of the cavity field. 
$\hat{b}^{+}$($\hat{b}$) is creation (annihilation)
operator of the exciton. 
The exciton and photon modes are coupled respectively to their reservoirs
with continuum modes which lead to 
dissipation.  $\hat{H}_{R_{1}}$ and $\hat{H}_{R_{2}}$ are  the free reservoir
 Hamiltonians. The reservoir coupling  may be  
considered as phonon scattering process in the case of exciton. But 
it is  a simple single
 photon absorbing process for the cavity field. 
$\Gamma_{2}^{+}(\Gamma_{2})$ and $\Gamma_{1}^{+}(\Gamma_{1})$ are the reservoir 
operators of the exciton and the cavity field with coupling constants included 
respectively. $G$ is the coupling constant of the effective exciton-exciton
interaction, it is a positive real number.

The  time evolution of the exciton and the cavity field system is described by
the master equation of the  reduced  density operator. In Schrodinger picture ,
the master equation, which is based on the Born-Markoff approximation, is 
written as following:
\begin{eqnarray}
\frac{\partial \hat{\rho}}{\partial t}
&=&\frac{1}{i\hbar}[\hat{H}_{0}, \hat{\rho}]
+\gamma_{1}(2\hat{a}\hat{\rho}\hat{a}^{+}-
\hat{\rho} \hat{a}^{+}\hat{a}
-\hat{a}^{+}\hat{a}\hat{\rho})   \nonumber \\
&+& \gamma_{2}
(2\hat{b}\hat{\rho}\hat{b}^{+}-\hat{\rho} \hat{b}^{+}\hat{b}
-\hat{b}^{+}\hat{b}\hat{\rho}), 
\end{eqnarray}
where $\gamma_{1}$ and $\gamma_{2}$  are dissipative rates of the cavity field 
 and the exciton.
\begin{center}
{\bf 3. Classical equation and small fluctuation analysis}
\end{center}
By using two modes positive P representation~\cite{B}, we can convert the 
operator equation (3) into c-number Fokker-Planck equation:
\begin{eqnarray}
\frac{\partial P(\alpha)}{\partial t} & = & \{
\frac{\partial}{\partial \alpha_{1}}
(\gamma_{1}\alpha_{1}-g\alpha_{2})
+\frac{\partial}{\partial \alpha^{+}_{1}}
(\gamma_{1}\alpha^{+}_{1}-g\alpha^{+}_{2})   \nonumber \\
&+&\frac{\partial}{\partial \alpha_{2}}
(\gamma_{2}\alpha_{2}-E +i2G\alpha^{+}_{2}\alpha^{2}_{2}
+g\alpha_{1})       \nonumber \\
&+&\frac{\partial}{\partial \alpha^{+}_{2}}
(\gamma_{2}\alpha^{+}_{2}-E^{*}+i2G\alpha_{2}\alpha^{+2}_{2}
+g\alpha^{+}_{1})   \nonumber \\
& - &i\frac{\partial^{2}}{\partial\alpha^{2}_{2}}G\alpha^{2}_{2}
+ i \frac{\partial^{2}}{\partial\alpha^{+2}_{2}}G\alpha^{+2}_{2}
\} P(\alpha),
\end{eqnarray}
where, $\alpha=
\left ( \begin{array}{c}
\alpha_{1} \\  \alpha^{+}_{1} \\  \alpha_{2} \\    \alpha^{+}_{2} 
\end {array} \right )$,     $\alpha_{1}$ is eigenvalue for the coherent state
of the light field and $\alpha_{2}$ is eigenvalue for the coherent state of 
the exciton. We have  transformed to the rotating frames of the cavity field 
and exciton by:
\begin{equation}
\left \{ \begin{array}{c}
\alpha_{1} \rightarrow \alpha_{1}e^{-i\omega_{c}t} \\
\alpha_{2} \rightarrow \alpha_{2}e^{-i\omega_{b}t}\\
\alpha^{+}_{1} \rightarrow \alpha^{+}_{1}e^{i\omega_{c}t} \\
\alpha^{+}_{2} \rightarrow \alpha^{+}_{2}e^{i\omega_{b}t}
\end{array} \right.
\end{equation}
In terms of the positive P representation, $\alpha_{1}$($\alpha_{2}$)
 and $\alpha^{+}_{1}$($\alpha^{+}_{2}$) are not complex conjugate to each 
 other.
The above Fokker-Planck equation (4) has a positive semi-definite  
 diffusion matrix leading a equivalent stochastic differential equation:
\begin{eqnarray}
\frac{\partial}{\partial t}
\left ( \begin{array}{c}
\alpha_{1} \\  \alpha^{+}_{1} \\  \alpha_{2} \\    \alpha^{+}_{2} 
\end{array} \right )
&=&-\left ( \begin{array}{l}
\gamma_{1}\alpha_{1}-g\alpha_{2} \\
\gamma_{1}\alpha^{+}_{1}-g\alpha^{+}_{2} \\
\gamma_{2}\alpha_{2}-E +i2G\alpha^{+}_{2}\alpha^{2}_{2}
+ g\alpha_{1}                 \\
\gamma_{2}\alpha^{+}_{2}-E^{*}-i2G\alpha_{2}\alpha^{+2}_{2}
+g\alpha^{+}_{1}
\end{array} \right )         \nonumber \\
&+& \left ( \begin{array}{cccc}
0 & 0 & 0 & 0 \\
0 & 0 & 0 & 0 \\
0 & 0 &-i2G\alpha^{2}_{2}  & 0 \\
0 & 0 & 0 & i2G\alpha^{+2}_{2}
\end{array} \right )^{\frac{1}{2}}         
\left ( \begin{array}{c}
\eta_{1}(t) \\  \eta^{+}_{1}(t) \\  \eta_{2}(t) \\    \eta^{+}_{2}(t) 
\end {array} \right ).
\end{eqnarray}
 $\eta_{i}(t)$ are stochastic fluctuation forces with zero means and satisfy
 the delta correlated function:
\begin{equation}
\left \{ \begin{array}{lcl}
<\eta_{i}(t)>&=&0 \\
<\eta^{+}_{i}(t)\eta_{j}(t^{\prime})>&=&0 \\
<\eta_{i}(t)\eta_{j}(t^{\prime})>&=&\delta_{ij}\delta(t-t^{\prime})\\
<\eta^{+}_{i}(t)\eta^{+}_{j}(t^{\prime})>&=&\delta_{ij}\delta(t-t^{\prime})
\end{array} \right.
.
\end{equation}
The exactly solutions of the non-linear equation (6) are still difficult to be 
obtained. Here we will use the small fluctuation approximation around the
stationary state to reduce the  non-linear equation (6) to a
linear equation for the fluctuations. A preliminary question is whether the
stationary state of the system is  stable against to small perturbation. 
To this end, we need neglect the 
fluctuation fores and first derive the stationary solution. Without
 the fluctuation force, equation (6) becomes:
\begin{eqnarray}
\frac{\partial}{\partial t}\alpha_{1}&=&-\gamma_{1}\alpha_{1}+g\alpha_{2}, \\
\frac{\partial}{\partial t}\alpha^{+}_{1}&=&-\gamma_{1}\alpha^{+}_{1}
+g\alpha^{+}_{2}, \\
\frac{\partial}{\partial t}\alpha_{2}&=&-\gamma_{2}\alpha_{2}+E -
i2G\alpha^{+}_{2}\alpha^{2}_{2}-g\alpha_{1},                 \\
\frac{\partial}{\partial t}\alpha^{+}_{2}&=&
-\gamma_{2}\alpha^{+}_{2}+E^{*}+i2G\alpha_{2}\alpha^{+2}_{2}
-g\alpha^{+}_{1}.
\end{eqnarray}
Now $\alpha_{1}$($\alpha_{2}$) and $\alpha^{+}_{1}$ 
($\alpha^{+}_{2}$) are complex conjugate as can be seen from the
equation themselves. 
The stationary solutions  $\alpha^{0}_{i}$ 
($\alpha^{+0}_{i}$) and $\alpha^{0}_{i}$ ($\alpha^{+0}_{i}$) are determined by:
\begin{eqnarray}
&&-\gamma_{1}\alpha^{0}_{1}+g\alpha^{0}_{2} =0 \\
&&-\gamma_{1}\alpha^{+0}_{1}+g\alpha^{+0}_{2}=0 \\
&&-\gamma_{2}\alpha^{0}_{2}+E -i2G\alpha^{+0}_{2}\alpha^{02}_{2}
-g\alpha^{0}_{1}=0 \\
&&-\gamma_{2}\alpha^{+0}_{2}+E^{*}+i2G\alpha^{0}_{2}\alpha^{+02}_{2}
-g\alpha^{+0}_{1}=0.
\end{eqnarray}
Then  we consider the case that the system deviates somewhat from its 
stationary state, setting:
\begin{equation}
\left \{ \begin{array}{c}
\alpha_{1}=\alpha^{0}_{1}+\delta\alpha_{1} \\
\alpha_{2}=\alpha^{0}_{2}+\delta\alpha_{2}\\
\alpha^{+}_{1}=\alpha^{+0}_{1}+\delta\alpha^{+}_{1} \\
\alpha^{+}_{2}=\alpha^{+0}_{2}+\delta\alpha^{+}_{2}
\end{array} \right. .
\end{equation}
Substitute the equation (16) into the equations (8-11) and only keep the 
fluctuations in the first order, we get the linear  motion equations for 
 $\delta\alpha_{i}$ and  $\delta\alpha^{+}_{i}$ as following:
\begin{equation}
\frac{\partial}{\partial t}\delta\alpha = -A \delta\alpha ,
\end{equation}
Where, $n_{2}=\alpha^{+0}_{2}\alpha^{0}_{2}$,   
$ \delta\alpha=
\left ( \begin{array}{c}
\delta\alpha_{1} \\  \delta\alpha^{+}_{1}\\
\delta\alpha_{2}\\    \delta\alpha^{+}_{2}
\end{array} \right ) $, $ A=
\left ( \begin{array}{cccc}
\gamma_{1} & 0 & -g & 0\\
0 & \gamma_{1} & 0 & -g \\
g & 0 & \gamma_{2}+i4Gn_{2}
 & i2G\alpha^{02}_{2}\\
0 & g & -i2G\alpha^{+02}_{2} & \gamma_{2}-i4Gn_{2}
\end{array} \right )$ .
Now we seek for the
solutions of the equation (17) of the form $e^{\lambda t}$.
 The eigenvalue $\lambda$ can be obtained from the equation 
\begin{equation}
|A-\lambda I|=0
.
\end{equation}
in which $I$ is identity matrix. The eigenvalue equation deduced from
 the equation (18) is
\begin{equation}
[(\gamma_{1}-\lambda)(\gamma_{2}-\lambda)+g^{2}]^{2}+
12G^{2}n^{2}_{2}(\gamma_{1}-\lambda)^{2}=0.
\end{equation}
The eigenvalues are :
\begin{eqnarray}
\lambda_{1,2}=\frac{1}{2} \{[\gamma_{1}+\gamma_{2}+i2\sqrt{3}Gn_{2}]
\pm \sqrt{(-\gamma_{1}+\gamma_{2}+i2\sqrt{3}Gn_{2})^{2}-4g^{2}} \}, \\
\lambda_{3,4}=\frac{1}{2} \{[\gamma_{1}+\gamma_{2}-i2\sqrt{3}Gn_{2}  ]
\pm \sqrt{(-\gamma_{1}+\gamma_{2}-i2\sqrt{3}Gn_{2})^{2}-4g^{2}} \}. 
\end{eqnarray}
When all the four eigenvalues have  the positive real parts, the stationary 
state ($\alpha_{i}^{0}$,$\alpha_{i}^{+0}$) is stable. That is, when any small
deviation develops from $\alpha_{i}^{0}$ and $\alpha_{i}^{+0}$, 
$\alpha_{i}$ and $\alpha_{i}^{+}$ will return to $\alpha_{i}^{0}$ and 
$\alpha_{i}^{+0}$. If the values of parameters (G, g $\gamma_{1}$, $\gamma_{2}$)
are in the range to make the absolution value of the real part of 
$\sqrt{(-\gamma_{1}+\gamma_{2} \pm i2\sqrt{3}Gn_{2})^{2}-4g^{2}}$ larger than
$\gamma_{1}+\gamma_{2}$, then $\alpha_{i}^{0}(\alpha_{i}^{+0})$ are stable
solutions. Now we consider the case of the stationary atate.
 we have linear equations for  $\delta\alpha_{i}$ and $\delta\alpha_{i}^{+}$:
\begin{eqnarray}
\frac{\partial}{\partial t}
\left ( \begin{array}{c}
\delta\alpha_{1} \\  \delta\alpha^{+}_{1}\\
\delta\alpha_{2}\\    \delta\alpha^{+}_{2}
\end{array} \right )
&=& -\left ( \begin{array}{cccc}
\gamma_{1} & 0 & -\Omega_{0} & 0\\
0 & \gamma_{1} & 0 & -\Omega_{0} \\
\Omega_{0} & 0 & \gamma_{2}+i4G\alpha^{0}_{2}\alpha^{+0}_{2}
 & i2G\alpha^{02}_{1}\\
0 & \Omega_{0} & -i2G\alpha^{+02}_{1} & \gamma_{2}
-i4G\alpha^{0}_{2}\alpha^{+0}_{2}
\end{array} \right )
\left ( \begin{array}{c}
\delta\alpha_{1} \\  \delta\alpha^{+}_{1}\\
\delta\alpha_{2}\\    \delta\alpha^{+}_{2}
\end{array} \right )       \nonumber \\
&+& \left ( \begin{array}{cccc}
0 & 0 & 0 & 0 \\
0 & 0 & 0 & 0 \\
0 & 0 &-i2G\alpha^{02}_{2}  & 0 \\
0 & 0 & 0 & i2G\alpha^{+02}_{2}
\end{array} \right )^{\frac{1}{2}}         
\left ( \begin{array}{c}
\eta_{1}(t) \\  \eta^{+}_{1}(t) \\  \eta_{2}(t) \\    \eta^{+}_{2}(t) 
\end {array} \right ) .
\end{eqnarray}

We abbreviate this equation as following:
\begin{equation}
\frac{\partial}{\partial t}\delta \vec{\alpha}(t)=-A\vec{\alpha}(t)+
D^{\frac{1}{2}}\vec{\eta}(t) ,
\end{equation}
and will discuss the squeezing properties of the output field  based on it in 
the next section.

\begin{center}
{\bf 4.Quadrature component squeezing of the output field}
\end{center}
 The fluctuation spectra of the cavity-field in the stationary state 
are the Fourier transform of the correlation functions 
$<\alpha_{i}(t+\tau),\alpha_{j}(t)>$ in the positive P representation:
\begin{equation}
S_{ij}(\omega)=\int_{-\infty}^{+\infty}e^{-i\omega\tau}
<\alpha_{i}(t+\tau),\alpha_{j}(t)>{\rm d}\tau ,
\end{equation}
where $<\alpha_{i}(t+\tau), \alpha_{j}(t)>=<\alpha_{i}(t+\tau)\alpha_{j}(t)>
-<\alpha_{i}(t+\tau)><\alpha_{j}(t)>$.
Substituting the equation (16) into the equation (24), we have:
\begin{equation}
S_{ij}(\omega)=\int_{-\infty}^{+\infty}e^{-i\omega\tau}
<\delta\alpha_{i}(t+\tau)\delta\alpha_{j}(t)>{\rm d}\tau .
\end{equation}
The fluctuation spectra will be obtained~\cite{C} as following:
\begin{equation}
S(\omega)=(A+i \omega I)^{-1}D(A^{T}-i \omega I)^{-1},
\end{equation}
where $I$ is the identity matrix and  $T$ stands for transpose. $A$ and $D$ are
defined by equation(23). After somewhat tedious calculation, the fluctuation spectrum
of the cavity field  will be obtained:
\begin{eqnarray}
S_{11}(\omega)&=& -i\frac{2G\alpha^{02}_{2}g^{2}}{\Lambda}
\{ 4G^{2}n^{2}_{2}(\gamma_{1}^{2}+\omega^{2})+
  [g^{2}+(\gamma_{1}+i\omega)
  (\gamma_{2}+i\omega-i4Gn_{2})]  \nonumber \\
&\times &[g^{2}+(\gamma_{1}-i\omega)
     (\gamma_{2}-i\omega-i4Gn_{2})]   \},  \\
S_{12}(\omega)&=&S_{21}(\omega)=\frac{8G^{2}g^{2}n^{2}_{2}}{\Lambda}
[g^{2}\gamma_{1}+\gamma^{2}_{1}\gamma_{2}+\omega^{2}\gamma_{2}],  \\
S_{22}(\omega)&=& i\frac{2G\alpha^{+02}_{2}g^{2}}{\Lambda}
\{ 4G^{2}n^{2}_{2}(\gamma_{1}^{2}+\omega^{2})+
  [g^{2}+(\gamma_{1}+i\omega)
  (\gamma_{2}+i\omega+i4Gn_{2})]  \nonumber \\
&\times &[g^{2}+(\gamma_{1}-i\omega)
     (\gamma_{2}-i\omega+i4Gn_{2})]   \} , 
\end{eqnarray}
where
\begin{equation}
\Lambda=|[(\gamma_{1}+i\omega)(\gamma_{2}+i\omega)+g^{2}]^{2}
+12G^{2}n^{2}_{2}(\gamma_{1}+i\omega)^{2}|^{2}.
\end{equation}

The creation and annihilation operators for output field 
$\hat{a}_{out}(t)$ and $\hat{a}^{+}_{out}(t)$ will be presented
by two quadrature phase components$\hat{X}_{\pm}$ as following:
\begin{equation}
\left \{ \begin{array}{l}
\hat{a}_{out}(t)=\frac{1}{2}[\hat{X}_{+}^{out}(t)+i\hat{X}_{-}^{out}(t)]
e^{i(\theta-\Omega t)} \\
\hat{a}^{+}_{out}(t)=\frac{1}{2}[\hat{X}_{+}^{out}(t)-i\hat{X}_{-}^{out}(t)]
e^{-i(\theta-\Omega t)} \\
\end{array} \right.
\end{equation} 
$\Omega$ is the frequency of the cavity field and $\theta$ is a reference phase.
 The normally-ordered  squeezing spectrum for each  quadrature 
phase component of the output field may be expressed by the fluctuation spectra
of the cavity fielde~\cite{C}:
\begin{eqnarray}
:S_{+}^{out}(\omega):&=&2\gamma_{1}[e^{-2i\theta}S_{11}(\omega)+
e^{2i\theta}S_{22}(\omega)+S_{21}(\omega)+S_{12}(\omega)], \\
:S_{-}^{out}(\omega):&=&2\gamma_{1}[-e^{-2i\theta}S_{11}(\omega)-
e^{2i\theta}S_{22}(\omega)+S_{21}(\omega)+S_{12}(\omega)].
\end{eqnarray}
If we choose the phase 
\begin{equation}
\theta=\theta_{1}-\frac{\pi}{4},
\end{equation}
where $\theta_{1}$ is the phase of the coherent state of the light field.  
The normal squeezing spectra of the quadrature components$\hat{X}_{\pm}$ for the 
output field are:
\begin{eqnarray}
:S^{\rm out}_{+}(\omega):&=&\frac{8Gg^{2}\gamma_{1}}{\Lambda}
[(g^{2}+\gamma_{1}\gamma_{2})^{2}+\omega^{4}+2Gn^{2}_{2}(g^{2}\gamma_{1}
+\gamma^{2}_{1}\gamma_{2})  \nonumber  \\
&+& (\gamma^{2}_{1}+\gamma^{2}_{2}-2g^{2}-12G^{2}n^{2}_{2}+
2G\gamma_{2}n^{2}_{2})\omega^{2}-12G^{2}n_{2}^{2}\gamma^{2}_{1}],\\
:S^{\rm out}_{-}(\omega):&=&\frac{8Gg^{2}\gamma_{1}}{\Lambda}
[-(g^{2}+\gamma_{1}\gamma_{2})^{2}-\omega^{4}+2Gn^{2}_{2}(g^{2}\gamma_{1}
+\gamma^{2}_{1}\gamma_{2})  \nonumber  \\
&-& (\gamma^{2}_{1}+\gamma^{2}_{2}-2g^{2}-12G^{2}n^{2}_{2}
-2G\gamma_{2}n^{2}_{2})\omega^{2}+12G^{2}n_{2}^{2}\gamma^{2}_{1}].
\end{eqnarray}
The most interesting part is the low frequency part. At $\omega=0$, 
the corresponding values are :
Now we consider  the case  of $\omega=0$ , it corresponds to the resonance 
output. That is: 
\begin{eqnarray}
:S^{\rm out}_{+}(0):&=&\frac{8G\gamma_{1}g^{2}}{\Lambda_{0}}
[(g^{2}+\gamma_{1}\gamma_{2})^{2}+2Gn_{2}^{2}(g^{2}\gamma_{1}
+\gamma_{1}^{2}\gamma_{2})-12G^{2}n_{2}^{2}\gamma^{2}_{1}], \\
:S^{\rm out}_{-}(0):&=&\frac{8G\gamma_{1}g^{2}}{\Lambda_{0}}
[-(g^{2}+\gamma_{1}\gamma_{2})^{2}+2Gn_{2}^{2}(g^{2}\gamma_{1}
+\gamma_{1}^{2}\gamma_{2})+12G^{2}n_{2}^{2}\gamma^{2}_{1}],
\end{eqnarray}
in which $\Lambda_{0}$ denotes $\Lambda(\omega=0)$. In the case
\begin{equation}
(g^{2}+\gamma_{1}\gamma_{2})^{2}+2Gn_{2}^{2}(g^{2}\gamma_{1}
+\gamma_{1}^{2}\gamma_{2})<12G^{2}n_{2}^{2}\gamma^{2}_{1},
\end{equation}
the $\hat{X}_{+}$ component will be squeezed at $\omega=0$.
If if
\begin{equation}
2Gn_{2}^{2}(g^{2}\gamma_{1}
+\gamma_{1}^{2}\gamma_{2})+12G^{2}n_{2}^{2}\gamma^{2}_{1}<
(g^{2}+\gamma_{1}\gamma_{2})^{2}
,
\end{equation}
the $\hat{X}_{-}$ component  will be squeezed at $\omega=0$.
It 
deserves to be mentioned that  when $G=0$, namely the interaction 
between the excitons can be neglected, the fluctuation spectra of the output
field are zero. This means that in this case the output field is a coherent 
light field as expected.  
\begin{center}
{\bf 5. Conclusion}
\end{center}
The quantum statistical properties of the field emitted by the high density 
exciton laser have been discussed. The conditions for the stability of 
 the stationary state  also have been given.
The small fluctuation approximation around the stationary state is 
made to deduce the fluctuation spectra.
We show that either the output  qudrature phase components  could be squeezed 
under certain conditions. 

\begin{center}
 {\bf Acknowledgments}
 \end{center}
One of authors (Liu Yu-Xi) wishes to express his gratitude to 
 Professor Sun Chang-Pu for  help in his work. This work is supported by
 National Natural Science Foundation of china, Grant number 19677204.

\end{document}